\documentclass{elsart}
\usepackage{latexsym}

\newcommand{\ZTOP}{\texttt{ZTOP}}
\newcommand{\FORTRAN}{\texttt{FORTRAN}}
\newcommand{\HERWIG}{\texttt{HERWIG}}
\newcommand{\PYTHIA}{\texttt{PYTHIA}}
\newcommand{\POLINT}{\texttt{POLINT}}
\newcommand{\POLINTT}{\texttt{POLINT3}}
\newcommand{\POLINTF}{\texttt{POLINT4}}
\newcommand{\POLINTTF}{\texttt{POLINT3/4}}
\newcommand{\STRUCTM}{\texttt{STRUCTM}}
\newcommand{\PFTOPDG}{\texttt{PFTOPDG}}
\newcommand{\ifc}{\texttt{ifc}}
\newcommand{\gnu}{\texttt{g77}}

\begin{document}

\date{13 December 2004}

\begin{frontmatter}

\title{Fast Evaluation of CTEQ Parton Distributions in Monte Carlos}

\author{Zack Sullivan}
\address{Theoretical Physics Department, Fermi National Accelerator
Laboratory, Batavia, IL 60510-0500, USA}

\begin{abstract}
A few changes to the routines that calculate CTEQ parton
distribution functions allow modern compilers to optimize the
evaluations, while having no quantitative effect on the results.
Overall computation time is reduced by a factor of 4--5 in
matrix-element calculations, and by 1.3--2.5 in showering Monte Carlo
event generators.  Similar improvements in performance may be expected
in any calculations relying heavily on interpolation or multiple calls
to functions.
\end{abstract}


\end{frontmatter}

\section{Introduction}
\label{sec:introduction}

A significant amount of time and computing resources are spent on
calculating events at hadron colliders.  Whether a theoretical
calculation of matrix elements, or an experimental simulation of
events with detector effects, one common element is the evaluation of
parton distribution functions (PDFs).  These functions return the
probability of finding a parton (quark or gluon) inside of a proton,
based on two parameters: the fraction of momentum carried by the
parton $x$, and the square of the energy scale of the process $Q^2$.
Because the input parameters can span several orders of magnitude, the
values of these functions are stored in two-dimensional tables for a
finite number of input points.  An approximate result for an arbitrary
input of $x$ and $Q^2$ is derived by interpolating between the values
obtained from the nearest table entries.

In profiling \ZTOP\ \cite{Harris:2002md,Sullivan:2004ie}, a \FORTRAN\
code written to simulate next-to-leading-order jet distributions in
single-top-quark production, it has become apparent that much of the
execution time of real production code is spent acquiring PDFs.  Upon
close examination of the CTEQ4 and CTEQ5 PDF codes \cite{CTEQ}, a
handful of trivial optimizations arise that can cut this time in half.
Based on this success, I examine further algorithmic improvements in
the typical interface functions that reduce the execution time by
another factor of two or more for all CTEQ PDFs (including CTEQ6
\cite{CTEQ6}).  I provide specific recommendations that are simple to
implement, but which can have large consequences for efficiency.

A \texttt{gprof} profile of \ZTOP\ \cite{Harris:2002md,Sullivan:2004ie}
indicates that up to 90\% of the execution time is spent in acquiring
PDFs.  Execution times in other programs appear to be dominated by the
same routines.  In Table~\ref{tab:CTEQdef} I show the typical fraction
of time spent evaluating PDFs for \ZTOP, and the two most common
showering event generators, \HERWIG\ 6.1 \cite{Corcella:2000bw}, and
\PYTHIA\ 6.2 \cite{Sjostrand:2000wi}.  The results in
Table~\ref{tab:CTEQdef} were generated using the GNU \gnu\ 3.1
compiler for linux on a 1.4 GHz Pentium 4 processor with the flags
\texttt{-g -pg -O3 -march=pentium4 -msse2}, and were verified
by commenting out the routines.  The results vary by less than 3\%
when changing compilers or compiler flags.
Times for CTEQ4 and CTEQ5 differ from CTEQ6, because the latter uses a
different interpolation algorithm.  Retrieving PDFs is always the most
time-intensive operation in these calculations.  Therefore, it
behooves us investigate what options are available to speed up the PDF
routines.

\begin{table}
\caption{Fraction of time spent evaluating PDF functions using default
CTEQ computer codes for three programs: the next-to-leading-order jet
calculation \ZTOP, and two showering event generators, \HERWIG\ and
\PYTHIA.\label{tab:CTEQdef}}
\medskip
\begin{center}
\begin{tabular}{ccc}\hline
Program&CTEQ4/5 &CTEQ6\\\hline
\ZTOP & 90\% & 60\% \\
\HERWIG & 70\% & 33\% \\
\PYTHIA & 35\% & 16\% \\\hline
\end{tabular}
\end{center}
\end{table}

Having identified the PDFs as the main bottleneck in the calculation
of cross sections and events at hadron colliders, I will examine
several successive levels of optimization in Sec.~\ref{sec:levels}.
The trade off will be that each level involves replacing a larger
fraction of the base code.  In Sec.~\ref{sec:results} I evaluate the
effectiveness of each change using a benchmark program, and three real
production codes: \ZTOP, \HERWIG, and \PYTHIA.  I conclude with some
observations and recommendations for future improvements.

\section{Levels of optimization}
\label{sec:levels}

It is important to recognize that code which evaluates PDFs is often
embedded in complex ways inside an application.  Hence, the
replacement of a given routine could be considered invasive.  I
consider classes of optimization that each replace larger portions of
code.  In practice the first two changes I suggest are easy to
accommodate.  However, the final one replaces a routine from a
commonly used library, and hence care must be taken to ensure that no
hidden dependencies arise.  In all cases the routines have been
verified to work with all programs mentioned in this paper.

\subsection{Modifying \POLINT}
\label{sec:polint}

Looking more carefully at a profile of \ZTOP\ using CTEQ5 PDFs
indicates that more than 75\% of the time is spent inside the
subroutine \POLINT.  \POLINT\ is a routine designed to perform a
polynomial fit of degree $n-1$ to a data set of $n$ points based on
\emph{Neville's algorithm} \cite{numrec}.  This subroutine is used by
the CTEQ Collaboration \cite{CTEQ,CTEQ6} to interpolate smoothly
between the values of $x$ and $Q^2$ that are read in from a table of
best-fit values.

One approach to increasing speed would be to replace \POLINT\ outright
with alternate interpolations, or functional fits to the PDFs.  While
these are reasonable choices, it is important to ensure that any
results are numerically identical to results obtained previously.
Therefore, I begin by making trivial modifications of \POLINT\ itself.
Two useful changes \cite{Sullivan:2004tt} are:
\begin{enumerate}
\item Remove the line: \texttt{IF(DEN.EQ.0.)PAUSE}.
\item Write different versions of \POLINT\ for 3- and 4-point interpolation,
and call them directly.  E.g., replace \texttt{POLINT(XA,YA,N,X,Y,DY)}
with\break \texttt{POLINT3(XA,YA,3,X,Y,DY)} in \texttt{PARTONX4} or
\texttt{PARTONX5}.
\end{enumerate}
The first optimization is the most important as the line is never
reached in the evaluation of the CTEQ PDFs, but it generally prevents
the compiler from fully optimizing the loops \cite{intelao}.  Beyond
being an unnecessary comparison, allowing a break point out of the
loop forces the processor to flush the instruction pipeline, and can
produce a missed branch comparison.  It also prevents the compiler
from using most types of parallel instructions.  The second
optimization mostly helps compilers to optimize the loops by defining
the number of iterations at compile time, rather than dynamically.

This optimization is more often effective with CTEQ4/5 than with
CTEQ6, since \POLINT\ is only used at the edges of the $x$ and $Q^2$
tables in CTEQ6.  However, using these optimizations with CTEQ6 will
never be slower, and can be much faster for some calculations.  We
will see in Sec.~\ref{sec:results} that using \POLINTT\ and \POLINTF\
has other benefits as well.

A further optimization for CTEQ4/5 comes from completely recoding
\emph{Neville's algorithm} for the special case of 3 points, and
removing the return of the error estimation.  As a general routine,
\POLINT\ evaluates several expressions that are never used if there
are only 3 points.  Furthermore, the CTEQ code does not use the error
estimate provided in the general case.  Therefore, all unnecessary
calculations and assignments are removed.  This results in a reduction
of the number of machine instructions, data reads, and data writes by
a factor of 3.  Net effects on the overall speed of execution are
described in Sec.~\ref{sec:results}.  The ``fast'' version of
\POLINTT\ is listed in the Appendix.

\subsection{Modifying \texttt{PARTONXN}}
\label{sec:partonxn}

Most programs that use parton distribution functions access them
through interface routines, such as \STRUCTM\ and \PFTOPDG\
\cite{CERNLIB}.  The key feature of these routines is that they ask
for the density of all partons at once ($u_v$, $u_s$, $d_v$, $d_s$,
$s$, $c$, $b$, $g$).  Typically this is done by looping over the
routines that access the PDFs, where the values of $x$ and $Q^2$ are
fixed, but only the flavor of the parton changes.  This immediately
suggests an algorithmic improvement that should be applicable to all
types of parton distributions: save the values of $x$ and $Q^2$, and
the results of any functions applied to them, and bypass those
functions unless $x$ or $Q^2$ change.

This algorithmic improvement in the CTEQ PDFs involves minor edits to
the routines \texttt{PARTONXN}, where \texttt{N} is the number of the
CTEQ set.  The changes consist of adding a few \texttt{SAVE}
statements, and a test for whether $x$ or $Q^2$ has changed.  For
CTEQ4/5 add
\begin{verbatim}
      DOUBLE PRECISION XLAST, QLAST, QG
      INTEGER JX,JQ
      DATA XLAST, QLAST / -1D0, -1D0 /
      DATA JX, JQ / 0, 0 /
      SAVE XLAST,QLAST,QG,JX,JQ

      IF ((X.EQ.XLAST).AND.(Q.EQ.QLAST))
     &   GOTO 99
      XLAST=X
      QLAST=Q
\end{verbatim}
after the declaration statements, and add a statement label 99 to the
first line that involves the parton flavor:
\begin{verbatim}
99   IF (IPRTN .GE. 3) THEN
\end{verbatim}
The calls to \POLINT\ should also be changed to one of the versions of
\POLINTT\ mentioned in Sec.~\ref{sec:polint}.

The CTEQ6 PDFs use a completely different interpolation through the
table of $x$ and $Q^2$.  Most of the calculations in \texttt{PARTONX6}
are associated with this interpolation, and therefore there is a
greater potential gain by adding
\begin{verbatim}
      DOUBLE PRECISION X, Q
      INTEGER JX, JQ
      DATA X, Q / -1D0, -1D0 /
      DATA JX, JQ / 0, 0 /
      SAVE X, Q, JX, JQ, JLX, JLQ
      SAVE SS, CONST1, CONST2, CONST3, CONST4
      SAVE CONST5, CONST6
      SAVE SY2, SY3, S23, TT, T12, T13, T23
      SAVE T24, T34, TY2, TY3
      SAVE TMP1, TMP2, TDET

      IF ((XX.EQ.X).AND.(QQ.EQ.Q)) GOTO 99
\end{verbatim}
after the declaration statements, and adding the statement label 99 to
the same place as in CTEQ4/5.  The calls to \POLINT\ should also be
changed to \POLINTF, as mentioned in the Sec.~\ref{sec:polint}.

\subsection{Modifying \STRUCTM\ and \PFTOPDG}
\label{sec:structm}

All of the optimizations suggested so far consist of modifying the
CTEQ routines.  However, there can still be a significant overhead in
calling the CTEQ routines.  Since most programs access the PDFs by
calling \STRUCTM\ or \PFTOPDG, a final improvement would be to
completely replace these routines with versions specialized to the
CTEQ PDFs.  The idea is to incorporate \texttt{PARTONXN} directly into
\STRUCTM, and to remove any additional redundancies.  These routines
may be obtained from the author or from Ref.~\cite{zspdf}.

There are two options for optimizing \PFTOPDG.  The first option is to
write another copy of the code in \STRUCTM\ that returns the values in
a different format.  The second option, which is used by the \texttt{CERNLIB}
\texttt{PDFLIB} routines, is to simply call \STRUCTM.  The first option
is error-prone, and only provides a 2\% improvement in speed.  The
second option is actually an interface bug in \texttt{CERNLIB}
\PFTOPDG, since it advertises that it will separately return the PDFs for
all quarks and anti-quarks.  This is fine for CTEQ4--6, but newer PDFs
may not have the same value for $s$ and $\bar{s}$.  Therefore, a call
to \PFTOPDG\ may quietly give incorrect results in the future.  Care
must be taken to ensure that a given set of PDFs are consistently
accessed.

\section{Optimization results}
\label{sec:results}

In order to assess the usefulness of these optimizations, I evaluate
four programs.  I consider a loop over PDFs, and three calculations of
$t$-channel single-top-quark production: \ZTOP, an analytic
next-to-leading order calculation of jet distributions
\cite{Harris:2002md,Sullivan:2004ie}, \HERWIG\ 6.1 \cite{Corcella:2000bw},
and \PYTHIA\ 6.2 \cite{Sjostrand:2000wi}.  All calculations are
compiled with the GNU \FORTRAN\ compiler \gnu\ versions 2.95 and 3.1,
and the Intel compiler \ifc\ 6.0.  Previously \cite{Sullivan:2004tt},
I have investigated the effects of the first optimization on \HERWIG\
and \PYTHIA\ while including fast detector simulation \texttt{SHW}
\cite{Carena:2000yx}.  The routines added by \texttt{SHW} contribute
a few percent to the overall execution time.  Hence, in order to
more effectively isolate the effects of the PDFs and reduce dependency
on unnecessary libraries, I do not use \texttt{SHW} here, or write out
any data.  All numerical results are from execution on a 1.4 GHz
Pentium 4 machine.  Limited tests performed on Pentium 3 machines are
completely consistent with the results described below.

\subsection{Benchmark for PDFs}

The most naive test of potential speed gains comes from looping over
parton distribution functions.  I probe values of $x$ between
$10^{-5}$ and $0.98$, and fix the scale to be $Q^2=(x*1960 \mathrm{\
GeV})^2$.  These choices avoid any possibility of anomalous gains due
to fast memory access in the level 2 cache.  However, this may
underestimate the benefit of using \POLINTF\ in the CTEQ6
distributions for processes at the Large Hadron Collider at CERN,
where larger values of $Q^2$ will be typical.  In
Ref.~\cite{Sullivan:2004tt}, I called the CTEQ routines directly.  For
this comparison, I access the routines using a simple version of
\STRUCTM.  This is more representative of the typical use of the PDFs,
and allows a direct comparison of all optimization levels.

In Table~\ref{tab:bench} I show the relative speed gain for each
correction broken down by compiler and PDF set.  The numbers are
normalized to the results obtained using \POLINTTF.  This choice is
based on the observation in Ref.~\cite{Sullivan:2004tt} that the
execution times in typical codes are dependent on detailed choices of
compiler flags.  However, by using \POLINTTF, this dependence tends to
disappear.  Hence, by using the lowest level of optimization, not only
do the programs operate up to factors of 2 faster, the speeds become
more dependent on algorithms and less dependent on compiler details.

\begin{table}
\caption{Typical speed gains  compared to \POLINTTF\ when looping over
all partons, and $10^{-5} < x < 0.98$.  Each column is separately
normalized.  The fastest time for each is also
listed.\label{tab:bench}}
\medskip
\begin{center}
\begin{tabular}{lcccc}\hline
&\multicolumn{2}{c}{\gnu\ 3.1(2.95)}&\multicolumn{2}{c}{\ifc\ 6.0}\\
Optimization& CTEQ4/5 &CTEQ6&CTEQ4/5&CTEQ6\\\hline
Default CTEQ&1/(1.1--1.2)&1.0 &1/(1.5--2.7)&1/1.03\\
\POLINTTF &1.0 &1.0 &1.0 &1.0 \\
\POLINTT\ (fast)&1.5&--- &1.2 &--- \\
\texttt{SAVE} $x, Q^2$&1.26&2.6&1.12&2.4\\
\texttt{SAVE} $x, Q^2$ (fast)&2.3&--- &1.9&--- \\
fastest \STRUCTM &3.1&3.1&4.6&2.7\\\hline
fastest times &40 s& 50 s& 17 s&35 s\\\hline
\end{tabular}
\end{center}
\end{table}

Table~\ref{tab:bench} demonstrates the speed enhancement due to saving
common results between calls, as described in Sec.~\ref{sec:partonxn}.
This is a mild enhancement for CTEQ4/5, where the dominant subroutine
is \POLINTT, but is a significant improvement for CTEQ6.  For CTEQ4/5,
I also show the results of using a fully optimized version of
\POLINTT, described in Sec.~\ref{sec:polint}, and the total
improvement when combined with saving the values.  At this level, the
typical gain over the base \POLINTTF\ is an additional factor of 2.

There are several possible ways to combine \POLINT\ and
\texttt{PARTONXN} into \STRUCTM.  The line labeled ``fastest
\STRUCTM'' lists the speed gain over \POLINTTF\ using all suggested
improvements.  The net enhancement over the default CTEQ distributions
ranges from a factor of 3 to more than a factor of 12.  All benchmarks
are somewhat artificial, but this is a good indication of the upper
range of improvements we might expect.

The final line of Table~\ref{tab:bench} shows the fastest times
achieved for an arbitrary fixed number of loops.  The first
observation is that the \ifc\ compiler tends to be a factor of 1--3
times faster than the \gnu\ compilers.  (The difference between \gnu\
2.95 and \gnu\ 3.1 tends to be less than 5--10\%.)  The second
observation is that the fastest CTEQ4/5 is up to a factor of 2 faster
than CTEQ6.  This effect comes entirely from the difference in
interpolation algorithms.  We should therefore expect the fractions of
time spent calling PDFs listed in Table~\ref{tab:CTEQdef} to become
smaller, and CTEQ4/5 to use less time than CTEQ6.

\subsection{Matrix element codes and \ZTOP}
\label{sec:me}

Benchmarks can be misleading.  Therefore, I consider the effects of
the optimizations of Sec.~\ref{sec:levels} on a working production
code of single-top-quark production
\cite{Harris:2002md,Sullivan:2004ie} called \ZTOP.  The results listed
in Table~\ref{tab:ztop} show a remarkably similar gain to the
benchmark scenario except at the fastest times.  Again, the
replacement of \POLINT\ by \POLINTT\ tends to remove the dependence on
compiler flags.  By using the replacement for \STRUCTM, an additional
factor of 2 is typical.  The \ifc\ compiler pushes this to a factor of
3 on a Pentium 4 processor by adding vectorization.  The last line of
Table~\ref{tab:ztop} shows the typical result that the Intel compiler
is a factor of 1.5--2 faster than \gnu.

\begin{table}
\caption{Typical speed gains for the matrix-element code \ZTOP\ relative to
\POLINTTF.  Each column is separately normalized.  The fastest time for
each is also listed.\label{tab:ztop}}
\medskip
\begin{center}
\begin{tabular}{lcccc}\hline
&\multicolumn{2}{c}{\gnu\ 3.1(2.95)}&\multicolumn{2}{c}{\ifc\ 6.0}\\
Optimization& CTEQ4/5 &CTEQ6&CTEQ4/5&CTEQ6\\\hline
Default CTEQ& 1/1.2& 1.0 & 1/(1.6--2.2)& 1.0\\
\POLINTTF & 1.0 & 1.0 & 1.0 & 1.0 \\
\POLINTT\ (fast)& 1.3& --- & 1.25& --- \\
\texttt{SAVE} $x, Q^2$& 1.2& 2.0& 1.13& 2.0\\
\texttt{SAVE} $x, Q^2$ (fast)& 1.7& --- & 1.7& --- \\
fastest \STRUCTM & 1.9& 2.15& 1.9, 2.7& 2.15\\\hline
fastest times &86 s& 98 s& 60, 42 s&62 s\\\hline
\end{tabular}
\end{center}
\end{table}

Not all matrix element calculations use all PDFs.  In the case of
$t$-channel single-top-quark production, one leg in the matrix-element
diagrams has only an incoming $b$ or $\bar b$ quark, or a gluon $g$.
Additional time might be saved by eliminating any unnecessary calls to
the PDFs.  In practice this can be very difficult to achieve, e.g.,
\HERWIG\ and \PYTHIA\ use the \texttt{PDFLIB} \STRUCTM\ \cite{CERNLIB}
interface as an abstraction.  In order to use individual PDFs, they
would have to incorporate a new interface.  In the case of \ZTOP, the
execution time can be reduced by a factor of 1.5 from the base
\POLINTTF, but the improvement is less significant as additional
optimizations are used.  In general, if there is a clear way to
eliminate extraneous calls to PDFs when coding matrix elements, it
should be implemented.

\subsection{\HERWIG\ and \PYTHIA}
\label{sec:herwig}

Theoretical calculations are typically at the matrix-element or jet
level, but experimental and careful phenomenological studies generally
resort to using showering Monte Carlo event generators.  These codes
are significantly more complex, and we should expect to see less gain
in efficiency as additional time is spent in showering and detector
simulation.  In order to assess the impact on the two most common
event generators, \HERWIG\ 6.1 \cite{Corcella:2000bw} and \PYTHIA\ 6.2
\cite{Sjostrand:2000wi}, I use them to calculate $t$-channel
single-top-quark production, including all showering effects.

While evaluating PDFs is indeed the most time-intensive operation in 
\PYTHIA, Table~\ref{tab:CTEQdef} tells us that no optimization can
provide more than about a factor of 1.5 improvement in speed.  In
Table~\ref{tab:pythia} we see that the \ifc\ compiler achieves almost
the full factor of 1.5, while \gnu\ can attain a factor of 1.2.  Both
results are significantly faster than the parameterizations of CTEQ
PDFs built into \PYTHIA.  Since using table-based PDFs is also
inherently more accurate, there appears to be no reason to continue
producing parameterizations.

\begin{table}
\caption{Typical speed gains for \PYTHIA\ relative to \POLINTTF.  Each
column is separately normalized.  The fastest time for each is also
listed.\label{tab:pythia}}
\medskip
\begin{center}
\begin{tabular}{lcccc}\hline
&\multicolumn{2}{c}{\gnu\ 3.1(2.95)}&\multicolumn{2}{c}{\ifc\ 6.0}\\
Optimization& CTEQ4/5 & CTEQ6 & CTEQ4/5 & CTEQ6 \\\hline
Default CTEQ& 1/1.03& 1.0& 1/(1.15--1.25)& 1.0\\
\POLINTTF & 1.0 & 1.0 & 1.0 & 1.0 \\
\POLINTT\ (fast)& 1.06& --- & 1.05& --- \\
\texttt{SAVE} $x, Q^2$& 1.04& 1.13& 1.05& 1.13\\
\texttt{SAVE} $x, Q^2$ (fast)& 1.1& --- & 1.1& --- \\
fastest \STRUCTM & 1.14& 1.15& 1.18& 1.15\\ \hline
fastest times &55 s&58 s&42 s&43 s\\ \hline
\end{tabular}
\end{center}
\end{table}

The \HERWIG\ event generator spends almost as much time evaluating
PDFs as the matrix element example considered above.  This can be
traced to \HERWIG\ calling \STRUCTM\ 1100--1800 times for each event
requested.  Table~\ref{tab:herwig} shows the speed gains when using
the each level of optimization as before.  Remarkably, execution times
are reduced by a factor of 1.8--2.5 for CTEQ4/5, and by 1.4--1.7 for
CTEQ6.  The actual times are somewhat anomalous, however.  Unlike the
benchmark, \ZTOP, or \PYTHIA, \gnu\ 3.1 appears to produce faster code
than \ifc\ 6.0, and CTEQ6 is faster than CTEQ4/5.  This is fortuitous
for people using the \gnu\ compiler, but there is a catch.  Using any
compiler, and any PDF code (including the default CTEQ PDFs), the
number of events produced by \HERWIG\ 6.1 differs by up to 5\% when
different compiler flags are used.  There appears to be a large
sensitivity to round-off errors when using initial-state showering.
This should be further investigated to determine whether this issue
affects all \HERWIG\ results, or is isolated to certain processes.

\begin{table}
\caption{Typical speed gains for \HERWIG\ relative to \POLINTTF.  Each
column is separately normalized.  The fastest time for each is also
listed.\label{tab:herwig}}
\medskip
\begin{center}
\begin{tabular}{lcccc}\hline
&\multicolumn{2}{c}{\gnu\ 3.1(2.95)}&\multicolumn{2}{c}{\ifc\ 6.0}\\
Optimization & CTEQ4/5 & CTEQ6 & CTEQ4/5 & CTEQ6 \\\hline
Default CTEQ& 1/1.12& 1.0& 1/(1.2--1.7)& 1.0\\
\POLINTTF & 1.0 & 1.0 & 1.0 & 1.0 \\
\POLINTT\ (fast)& 1.25& --- & 1.1& --- \\
\texttt{SAVE} $x, Q^2$& 1.14& 1.6& 1.12& 1.3\\
\texttt{SAVE} $x, Q^2$ (fast)& 1.5& --- & 1.3& --- \\
fastest \STRUCTM & 1.65& 1.7& 1.53& 1.4\\ \hline
fastest times &64 s&50 s&75 s&55 s\\ \hline
\end{tabular}
\end{center}
\end{table}

\section{Conclusions}

Given the increasingly complex nature of calculations of hadronic
physics we should investigate where bottlenecks in computational speed
arise.  It appears that one source of significant loss of
computational speed is in evaluating parton distribution functions.
For users of the CTEQ PDFs I propose three levels of optimization that
can reduce computational times by up to a factor of 1.3--2.5 in
showing event generators, and up to a factor of 4--5 in matrix-element
calculations.

\begin{table}
\caption{Fraction of time spent evaluating PDF functions using an
optimized \STRUCTM\ for three programs: the next-to-leading-order jet
calculation \ZTOP, and two showering event generators, \HERWIG\ and
\PYTHIA.
\label{tab:fraction}}
\medskip
\begin{center}
\begin{tabular}{ccc}\hline
Program&CTEQ4/5 &CTEQ6\\\hline
\ZTOP & 42\% & 48\% \\
\HERWIG & 30\% & 23\% \\
\PYTHIA & 9\% & 9\% \\\hline
\end{tabular}
\end{center}
\end{table}

We observe in Sec.~\ref{sec:results} that replacing \POLINT\ with
\POLINTTF\ greatly reduces the dependence of program execution time on
the choice of compiler flags.  This is a simple change, and can
improve running times of some programs by up to a factor of 2.  Since
most programs call several PDFs in a row with the same values of $x$
and $Q^2$, the obvious next step is to modify the CTEQ routines
\texttt{PARTONXN} to save previous results, and calculate only what
has changed.  The third level of optimization replaces the typical
\texttt{CERNLIB} interface functions \STRUCTM\ and \PFTOPDG\
\cite{CERNLIB} with fully optimized versions that are specialized to
the CTEQ parton distributions.  I recommend that these improvements be
incorporated into the base CTEQ distribution \cite{CTEQ}, the
\texttt{PDFLIB} routines in \texttt{CERNLIB} \cite{CERNLIB}, and the
new Les Houches Accord compilation of PDFs \texttt{LHAPDF}
\cite{Giele:2002hx}.  Full versions of the routines presented here may
be obtained from the author, or from Ref.~\cite{zspdf}.

Despite these impressive gains, we see in Table~\ref{tab:fraction}
that evaluating PDFs remains the most time-consuming aspect of
hadronic calculations.  This suggests two avenues of investigation
that should be considered for future programs.  First, a systematic
study of PDF evolution codes should be performed to determine
whether there are more efficient interpolation algorithms to use with
table-based PDFs.  This would allow universal improvements in code
execution.  Second, each Monte Carlo writer should be aware of the
timing issues (and potential bugs if $\bar s$ is not the same as $s$),
and consider using an interface structure other than \STRUCTM.  The
potential savings from calling one PDF instead of eight or more could
be very significant.  Finally, it is interesting that the
optimizations I have listed can remove the apparent need for
parameterizations of the parton distribution functions.
In general, any calculation that relies heavily on interpolation, or
multiple evaluations of a function in which some pieces do not vary,
should see similar improvements in performance by applying these same
techniques.

\begin{ack}
This work was supported by Universities Research Association
Inc.\ under Contract No.\ DE-AC02-76CH03000 with the United States
Department of Energy.
\end{ack}

\appendix*
\section{A fast \POLINTT}
\label{sec:polintt}

This is a ``fast'' version of \POLINTT, which has been optimized for
the special case of 3-point fitting, and no possibility of divisions
by zero.  An error estimate is not returned, since it is never used in
the CTEQ evolution codes.  This code should be used to replace the
version of \POLINT\ called by the CTEQ4 and CTEQ5 PDFs.

\begin{verbatim}
C  This is a specialized recoding of Neville's
C   algorithm based on the POLINT routine from
C   "Numerical Recipes", but assuming N=3, and
C   ignoring the error estimation.
C  Written by Z. Sullivan, May 2004
C  This file uses a minimal number of
C   instructions to do 3-point fitting.
C     SUBROUTINE POLINT  (XA,YA,3,X,Y,IGNORED)
      SUBROUTINE POLINT3 (XA,YA,N,X,Y,DY)
      IMPLICIT NONE
      DOUBLE PRECISION XA(3),YA(3),X,Y,DY,DEN
      DOUBLE PRECISION C1,HO,HP,HP2,W,D1,D2
      INTEGER N

      HO=XA(1)-X
      HP=XA(2)-X
      W=YA(2)-YA(1)
      DEN=HO-HP
      DEN=W/DEN
      D1=HP*DEN
      C1=HO*DEN

      HP2=XA(3)-X
      W=YA(3)-YA(2)
      DEN=HP-HP2
      DEN=W/DEN
      D2=HP2*DEN

      W=HP*DEN-D1
      DEN=HO-HP2

      IF((X+X-XA(2)-XA(3)).GT.0D0) THEN
         Y=YA(3)+D2+HP2*W/DEN
      ELSEIF((X+X-XA(1)-XA(2)).GT.0D0) THEN
         Y=YA(2)+D1+HO*W/DEN
      ELSE
         Y=YA(1)+C1+HO*W/DEN
      ENDIF

      RETURN
      END
\end{verbatim}

\end{document}